# Extending the class of solutions of the Dirac equation using tridiagonal matrix representations

A. D. Alhaidari[a], H. Bahlouli[a,b] and I. A. Assi[b]

[a] *Saudi Center for Theoretical Physics, P.O. Box 32741, Jeddah 21438, Saudi Arabia*
[b] *Physics Department, King Fahd University of Petroleum & Minerals, Dhahran 31261, Saudi Arabia*

**Abstract**: The aim of this work is to find exact solutions of the one-dimensional Dirac equation that do not belong to the already known conventional class. We write the spinor wavefunction as a bounded infinite sum in a complete basis set, which is chosen such that the matrix representation of the Dirac wave operator becomes tridiagonal and symmetric. This makes the wave equation equivalent to a symmetric three-term recursion relation for the expansion coefficients of the wavefunction. We solve the recursion relation and obtain the relativistic energy spectrum and corresponding state functions. Restriction to the diagonal representation results in the conventional class of solutions. As illustration, we consider the square well with half-sine potential bottom in the vector and scalar coupling under spin symmetry. We obtain the relativistic energy spectrum and show that the nonrelativistic limit agrees with results found elsewhere.

## 1. Introduction:

The Dirac equation is a relativistically covariant first order linear differential equation in space and time for multi-component wavefunction. It was formulated by Dirac in the early days of quantum theory to describe the dynamics of elementary spin ½ particles. The equation is invariant under Lorentz transformation as required by the theory of special relativity. Exact solutions of the Dirac equation are of great importance since they allow for a better understanding of the quantum behavior of spinor particles at high speeds or strong coupling to their environment. Moreover, such solutions allow for testing the validity of perturbative, variational and numerical approaches designed to solve the relativistic problem. In particular, the nonrelativistic limit of the solutions of the Dirac equation establishes the necessary correspondence between the relativistic effects and their nonrelativistic counterpart for a given physical configuration. However, it is worth mentioning that this correspondence is subtle as explained in our work [1]. For example, one may find that the nonrelativistic limit is unique but a relativistic extension of the nonrelativistic problem is not.

The number of exactly solvable potentials associated with the relativistic Dirac equation is very limited. In this context, exact solvability requires full knowledge of all eigenfunctions and the complete energy spectrum for a continuous range of values of the potential parameters. In contrast, conditionally exact solutions are generated only for specific values of the potential parameters and for quasi-exact solutions, only part of the energy spectrum is determined. Exact solutions of the Dirac equation in the presence of different potentials were obtained in the past [2-5] using various theoretical techniques.

Among these techniques, we mention the following few: (1) supersymmetry where one generalizes the concept of raising and lowering operators in the context of the harmonic oscillator [6], (2) Darboux transformation, which allows for a unified approach to isospectral potential problems [7], and (3) the intertwining and factorization approaches, which are essentially similar to the Darboux transformation [8].

In the present work, we construct a relativistic extension of our own approach for obtaining solvable potentials in quantum theory; the so-called tridiagonal representation approach [9]. We do that for the Dirac equation in 1+1 space-time. We start in the next section by setting up the mathematical formulation of the most general one-dimensional Dirac equation and then work out our strategy in obtaining a tridiagonal matrix representation for the wave operator, usually called the J-matrix operator. We have succeeded in obtaining the necessary conditions for solvability and implemented our approach to generate some solvable problems whose nonrelativistic limit agrees with results obtained in previous work. Details of our computations were relegated to two appendices, one devoted to the Jacobi basis set and the other to the Laguerre basis set. These two basis sets constitute the essential ingredients and are most often used in the tridiagonal representation approach.

## 2. Formulation:

In the relativistic units, $\hbar = c = 1$, the most general linear massive Dirac equation in 1+1 space-time dimension with time-independent potentials reads as follows:

$$\left\{\gamma^\mu\left[i\partial_\mu - A_\mu(x)\right] - S(x) - \gamma^5 W(x)\right\}\Psi(t,x) = m\Psi(t,x), \tag{1}$$

where $\left\{\gamma^\mu\right\}_{\mu=0,1}$ are the two Dirac gamma matrices such that $\left(\gamma^0\right)^2 = 1$, $\left(\gamma^1\right)^2 = -1$ and $\gamma^0\gamma^1 = -\gamma^1\gamma^0$. $S$ is the scalar potential, $A_\mu = (V, U)$ is the (time, space) component of the vector potential, $W$ is the pseudo-scalar potential, and $\gamma^5 = i\gamma^0\gamma^1$. Choosing the minimum dimensional representation for the gamma matrices defined by $\gamma^0 = \begin{pmatrix} 1 & 0 \\ 0 & -1 \end{pmatrix}$ and $\gamma^1 = -i\begin{pmatrix} 0 & 1 \\ 1 & 0 \end{pmatrix}$, gives $\gamma^5 = \begin{pmatrix} 0 & 1 \\ -1 & 0 \end{pmatrix}$ and makes the Dirac equation (1) look as follows in matrix form

$$\begin{pmatrix} m + S(x) + V(x) & -\frac{d}{dx} + W(x) - iU(x) \\ \frac{d}{dx} + W(x) + iU(x) & -m - S(x) + V(x) \end{pmatrix}\begin{pmatrix} \psi_+(x) \\ \psi_-(x) \end{pmatrix} = \varepsilon\begin{pmatrix} \psi_+(x) \\ \psi_-(x) \end{pmatrix}, \tag{2}$$

where we wrote the two-component spinor wavefunction as $\Psi(t,x) = e^{-i\varepsilon t}\psi(x)$. The space component of the vector potential, $U$, could be eliminated by a gauge transformation. Therefore, from now on we take $U = 0$. The objective now is to find a discrete spinor basis in which the wave equation (2) is represented by a tridiagonal symmetric matrix such that the corresponding three-term recursion relation could be solved exactly for as large a class of potentials as possible. It should be noted that in 2+1 space-time with cylindrical symmetry, Eq. (2) with $W(x) \to W(x) + \frac{\kappa}{x}$ represents the radial Dirac equation with $x$ being the radial coordinate and the angular quantum number $\kappa = \pm\frac{1}{2}, \pm\frac{3}{2}, \pm\frac{5}{2}, \ldots$

We expand the components of the spinor wavefunction in a complete basis set, $\{\phi_n^\pm\}_{n=0}^\infty$, as $|\psi^\pm(x)\rangle = \sum_n f_n(\varepsilon)|\phi_n^\pm(x)\rangle$, where $\{f_n\}_{n=0}^\infty$ are the expansion coefficients. The basis functions contain only kinematic information, which is common to all systems that share the same space-time configuration. However, structural and dynamical information about a specific physical system is contained in the expansion coefficients that depend not only on the system's energy but also on the potential parameters. Now, the matrix representation of the wave operator (2) in this basis reads as follows

$$\begin{aligned} J_{n,m} &= \langle\phi_n|(H-\varepsilon)|\phi_m\rangle = \langle\phi_n^+|(V_+ + m - \varepsilon)|\phi_m^+\rangle + \langle\phi_n^-|(V_- - m - \varepsilon)|\phi_m^-\rangle \\ &+ \langle\phi_n^+|\left(-\tfrac{d}{dx}+W\right)|\phi_m^-\rangle + \langle\phi_n^-|\left(\tfrac{d}{dx}+W\right)|\phi_m^+\rangle \end{aligned} \tag{3}$$

where $V_\pm = V \pm S$. Next, we make a coordinate transformation $x \to y(x)$ and relate the lower and upper components of the spinor basis as required by the kinetic balance relation [10]

$$\phi_n^-(x) = \frac{1}{\lambda}\left[\frac{d}{dx} + R(x)\right]\phi_n^+(x) = \frac{1}{\lambda}\left[y'\frac{d}{dy} + R(y)\right]\phi_n^+(y), \tag{4}$$

where $\lambda$ is a real scale parameter, $R(y)$ is an arbitrary real function to be determined later, and $y' = \frac{dy}{dx}$. Integration by parts in the right side of Eq. (3) makes the last term equal to the third but with the index interchange $n \leftrightarrow m$. However, such an integration is permitted only if the product $\phi_n^+(x)\phi_m^-(x)$ vanishes at the boundaries of configuration space. We will confirm that by example in the following section as shown in Table 1 and displayed in the basis components plot of Fig. 1 where such boundary conditions are satisfied. Using Eq. (4), we can rewrite the third term in (3) as follows

$$\begin{aligned} &\langle\phi_n^+|\left(-\tfrac{d}{dx}+W\right)|\phi_m^-\rangle = \\ &\frac{1}{\lambda}\langle\phi_n^+|\left[-y'^2\frac{d^2}{dy^2} - \left(y'' + y'R - y'W\right)\frac{d}{dy} - y'\dot{R} + WR\right]|\phi_m^+\rangle, \end{aligned} \tag{5}$$

where the dot over a variable stands for its derivative with respect to $y$; that is, $\dot{R} = \frac{dR}{dy}$. On the other hand, the second term could be manipulated using integration by parts to read as follows

$$\begin{aligned} &\langle\phi_n^-|(V_- - m - \varepsilon)|\phi_m^-\rangle = \\ &\frac{1}{\lambda^2}\langle\phi_n^+|Q_-\left[-y'^2\frac{d^2}{dy^2} - \left(y'' + y'^2\frac{\dot{Q}_-}{Q_-}\right)\frac{d}{dy} - y'\left(\dot{R} + \frac{\dot{Q}_-}{Q_-}R\right) + R^2\right]|\phi_m^+\rangle, \end{aligned} \tag{6}$$

where $Q_\pm = V_\pm \pm m - \varepsilon$. With the $n \leftrightarrow m$ exchange symmetry and the above results, we can rewrite Eq. (3) as

$$\begin{aligned} J_{n,m} &= -\frac{1}{\lambda}\langle\phi_n^+|\left\{\left(1+\tfrac{Q_-}{2\lambda}\right)y'^2\frac{d^2}{dy^2} + \left[y''\left(1+\tfrac{Q_-}{2\lambda}\right) + y'\left(R - W + y'\tfrac{\dot{Q}_-}{2\lambda}\right)\right]\frac{d}{dy}\right\}|\phi_m^+\rangle \\ &+ \frac{1}{\lambda}\langle\phi_n^+|\left[\frac{\lambda Q_+}{2} + \left(1+\tfrac{Q_-}{2\lambda}\right)\left(R^2 - y'\dot{R}\right) - R\left(R - W + y'\tfrac{\dot{Q}_-}{2\lambda}\right)\right]|\phi_m^+\rangle + n \leftrightarrow m \end{aligned} \tag{7}$$

In what follows, we look for specific realizations of the potential functions $\{V,S,W\}$, coordinate transformation $y(x)$, basis component $\phi_n^+(x)$, and kinetic balance function $R(x)$ such that the matrix representation of the Dirac wave operator $J$ (7) becomes tridiagonal and symmetric. After choosing $y(x)$, our strategy for achieving the tridiagonal requirement is to start by writing the basis element $\phi_n^+(x)$ in terms of classical polynomials whose orthogonality interval coincides with the range of $y(x)$. Next, we impose on (7) the second order differential equation satisfied by these polynomials and their differential property. The result is an equation that relates $R(x)$ to the potential functions and a linear relationship among the basis parameters. Explicit details of such procedure will be given in Appendix A for the Jacobi basis and in Appendix B for the Laguerre basis.

### 3. The Jacobi Basis: Square well with half-sine potential bottom example

If we choose the coordinate transformation $y(x)$ such that $y \in [-1,+1]$, then a compatible spinor basis could be constructed with the following upper component†

$$\phi_n^+(y) = A_n (1-y)^{\alpha} (1+y)^{\beta} P_n^{(\mu,\nu)}(y), \tag{8}$$

where $A_n = \sqrt{\frac{2n+\mu+\nu+1}{2^{\mu+\nu+1}} \frac{\Gamma(n+1)\Gamma(n+\mu+\nu+1)}{\Gamma(n+\nu+1)\Gamma(n+\mu+1)}}$, $P_n^{(\mu,\nu)}(y)$ is the Jacobi polynomial of degree $n$ in $y$, and the real parameters $\{\alpha,\beta,\mu,\nu\}$ will be determined later. The lower component of the spinor basis $\phi_n^-(x)$ is obtained from the upper (8) using the kinetic balance relation (4). Now, the coordinate transformation $y(x)$ is chosen such that $y' = \lambda(1-y)^a(1+y)^b$, where $a$ and $b$ are real positive parameters. This choice makes $y'$ resemble the weight factor that appears in front of the Jacobi polynomial in (8) and in its orthogonality relation (A5), thus making the calculation manageable. For example, if we consider $y(x) = \tanh(\lambda x)$ where $x \in [-\infty,+\infty]$ then $a = b = 1$. In Appendix A, we use the differential equation of the Jacobi polynomial (A3) and thus impose condition (A7) on the problem parameters. Moreover, we use the differential property of these polynomials (A4) to relate $R(x)$ to the potential functions as shown in (A8). Consequently, the integral (7) reduces to

$$J_{n,m} = 2c\lambda\Omega_{n,m} + 2\lambda A_n A_m \int_{-1}^{+1} (1-y)^{\mu}(1+y)^{\nu} G(y) P_n^{(\mu,\nu)}(y) P_m^{(\mu,\nu)}(y)\, dy, \tag{9}$$

where $\Omega$ is the tridiagonal symmetric matrix whose elements are shown in (A11) and the function $G(y)$ is given by (A10). The details of this manipulation are given in the Appendix. Now, the orthogonality relation of the Jacobi polynomials (A5) and their recursion property (A6) show that the matrix representation for the Dirac wave operator $J$ is tridiagonal if and only if $G(y)$ is a linear function of $y$. That is, if and only if $G(y) = \rho y + \sigma$, where $\rho$ and $\sigma$ are two real constant parameters.

Beyond this point, one may proceed only after selecting special cases by starting with a specific coordinate transformation $y(x)$ or, equivalently, a choice of parameters $a$ and $b$. As illustrative example, we consider the coordinate transformation $y(x) = \sin(\pi x/L)$ where $x \in [-\frac{1}{2}L, +\frac{1}{2}L]$. This gives $a = b = 1/2$ and $\lambda = \pi/L$. Moreover, we take $S = V$ and $W=0$. Thus, $V_-=0$ and using the constraint (A8) gives $R=cy'=c\lambda\sqrt{1-y^2}$. Putting all of this together in Eq. (A10), we obtain

$$G(y)=\frac{2V+m-\varepsilon}{2\lambda}-c\left[\beta-\alpha-(\alpha+\beta)y+c\left(1-y^2\right)\right]+\left(1-\tfrac{m+\varepsilon}{2\lambda}\right)\Big\{\left(n+\tfrac{\mu+\nu+1}{2}\right)^2 -\frac{1}{4}\left(\mu^2+\nu^2-\frac{1}{2}\right)-\frac{1}{4}\left(\mu^2-\frac{1}{4}\right)\frac{1+y}{1-y}-\frac{1}{4}\left(\nu^2-\frac{1}{4}\right)\frac{1-y}{1+y}+c\left[c(1-y^2)+y\right]\Big\} \tag{10}$$

† Other possibilities include $y \ge 0$ with $\phi_n^+(y) = A_n y^{\alpha} e^{-\beta y} L_n^{\nu}(y)$ and $y' = \lambda y^a e^{by}$, where $L_n^{\nu}(y)$ is the Laguerre polynomial (see Appendix B), or $y \in [-\infty,+\infty]$ with $\phi_n^+(y) = A_n e^{-y^2/2} H_n(y)$, where $H_n(y)$ is the Hermite polynomial, etc.

Thus, we conclude that a tridiagonal representation for *energy independent* potentials is obtained for $c=0$, $\mu^2=\nu^2=\frac{1}{4}$ and with $V(x)=V_0y=V_0\sin\left(\pi x/L\right)$. Using the recursion relation and orthogonal property of the Jacobi polynomials, we obtain the following tridiagonal matrix representation of the Dirac wave operator

$$J_{n,m}=2\lambda\left[\frac{m-\varepsilon}{2\lambda}+\left(1-\frac{m+\varepsilon}{2\lambda}\right)\left(n+\tfrac{\mu+\nu+1}{2}\right)^2\right]\delta_{n,m}+V_0\left(\delta_{n,m+1}+\delta_{n,m-1}\right), \tag{11}$$

where $\mu=\nu=\pm\frac{1}{2}$ or $\mu=-\nu=\pm\frac{1}{2}$. Note that the diagonal representation is obtained only if $V_0=0$ giving the energy spectrum formula

$$\varepsilon_n=m+2(\lambda-m)\frac{\left(n+\frac{\mu+\nu+1}{2}\right)^2}{1+\left(n+\frac{\mu+\nu+1}{2}\right)^2}. \tag{12}$$

Taking the nonrelativistic limit, $\varepsilon\approx m+E$ where $E<<m$, in (11) we obtain

$$E_n=2\lambda\left(1-\tfrac{m}{\lambda}\right)\left(n+\tfrac{\mu+\nu+1}{2}\right)^2, \tag{13}$$

which is identical to the energy spectrum of the infinite square well with a flat bottom but with a modified energy scale λ. Agreement is achieved by the parameter map $\lambda\to m+\left(\lambda/2\right)^2$. Expanding the spinor wavefunction as $|\psi\rangle=\sum_n f_n|\phi_n\rangle$ and substituting in the wave equation, $H|\psi\rangle=\varepsilon|\psi\rangle$, we obtain with the use of (11) the following three-term recursion relation for the expansion coefficients $\{f_n\}$

$$\frac{\varepsilon-m}{2\lambda}f_n=\left(1-\frac{m+\varepsilon}{2\lambda}\right)a_nf_n+\frac{V_0}{2\lambda}\left(f_{n-1}+f_{n+1}\right), \tag{14}$$

where $a_n=\left(n+\frac{\mu+\nu+1}{2}\right)^2$. Equation (14) is the equivalent discrete wave equation for the system under study. To write the upper component of the spinor basis explicitly, we use the following representations of the Chebyshev polynomials of the first, second, third, and forth kind in terms of the Jacobi polynomials

$$P_n^{\left(-\frac{1}{2},-\frac{1}{2}\right)}(y)=\frac{\Gamma(n+1/2)}{\sqrt{\pi}\,\Gamma(n+1)}T_n(y),\quad P_n^{\left(+\frac{1}{2},+\frac{1}{2}\right)}(y)=\frac{2\Gamma(n+3/2)}{\sqrt{\pi}\,\Gamma(n+2)}U_n(y), \tag{15a}$$

$$P_n^{\left(-\frac{1}{2},+\frac{1}{2}\right)}(y)=\frac{\Gamma(n+1/2)}{\sqrt{\pi}\,\Gamma(n+1)}V_n(y),\quad P_n^{\left(+\frac{1}{2},-\frac{1}{2}\right)}(y)=\frac{\Gamma(n+1/2)}{\sqrt{\pi}\,\Gamma(n+1)}W_n(y), \tag{15b}$$

where $W_n(y)=(-1)^nV_n(-y)$. We can now write the lower component of the spinor basis by using Eq. (4) and employing the following differential properties of these Chebyshev polynomials [11]

$$\frac{d}{dy}T_n(y)=nU_{n-1}(y),\quad \frac{d}{dy}\left[\sqrt{1-y^2}U_n(y)\right]=-\frac{n+1}{\sqrt{1-y^2}}T_{n+1}(y), \tag{16a}$$

$$\frac{d}{dy}\left[\sqrt{1+y}V_n(y)\right]=\frac{n+1/2}{\sqrt{1+y}}W_n(y),\ \frac{d}{dy}\left[\sqrt{1-y}W_n(y)\right]=-\frac{n+1/2}{\sqrt{1-y}}V_n(y). \tag{16b}$$

Consequently, we obtain the upper and lower components of the spinor basis associated with the four cases $\mu=\nu=\pm\frac{1}{2}$ and $\mu=-\nu=\pm\frac{1}{2}$ as shown in Table 1. In Figure 1, we plot these for a given index *n* showing the oscillatory behavior and the typical (almost $\pi/2$) phase shift between the upper and lower components $\phi_n^\pm(x)$. It is evident that they satisfy the boundary condition that the product $\phi_n^+(x)\phi_m^-(x)$ vanishes at the boundaries of configuration space. Now, taking the nonrelativistic limit of (14), one obtains

$$E\,f_n=2\left(\lambda-m\right)a_nf_n+V_0\left(f_{n-1}+f_{n+1}\right). \tag{17}$$

For the case $\mu=\nu=+\frac{1}{2}$, we obtain

$$E\,f_n=2\left(\lambda-m\right)(n+1)^2f_n+V_0\left(f_{n-1}+f_{n+1}\right). \tag{18}$$

This should agree with our findings in [12] for the half-sine potential bottom case that corresponds to $V(x)=\frac{1}{2}\lambda^2C\cos(\lambda x)$ with $x\in[0,+L]$ and, where the constant *C* in [12] corresponds to $2V_0/\lambda^2$. That result (in the atomic units, $\hbar=m=1$) reads as follows

$$E\,f_n=\tfrac{1}{2}\lambda^2(n+1)^2f_n+\tfrac{1}{2}V_0\left(f_{n-1}+f_{n+1}\right). \tag{19}$$

The parameters map $\lambda\to m+\left(\lambda/2\right)^2$ and $V_0\to\frac{1}{2}V_0$ makes the recursions (18) identical to (19). The first part of the map has already been established above below the energy spectrum formula (13). The second could be understood by noting that in the relativistic theory we took $S=V$, which in effect doubles the potential strength $V_0$ in (18). This potential strength doubling is explained in [13].

To obtain an exact solution of the relativistic recursion relation (14), we start by introducing the energy variable $\mathcal{E}$ defined by $\mathcal{E}=\varepsilon-m$. In terms of this energy variable, Eq. (14) becomes

$$\frac{\mathcal{E}}{2\lambda}\left(1+a_n\right)f_n=\left(1-\frac{m}{\lambda}\right)a_nf_n+\frac{V_0}{2\lambda}\left(f_{n-1}+f_{n+1}\right). \tag{20}$$

Defining $g_n=\sqrt{\frac{1+a_n}{1+a_0}}f_n$, we can rewrite this as a three-term recursion relation for $g_n$ as follows

$$\frac{\mathcal{E}}{2\lambda}g_n(\mathcal{E})=\left(1-\frac{m}{\lambda}\right)\mathcal{A}_ng_n(\mathcal{E})+\frac{V_0}{2\lambda}\left[\mathcal{B}_{n-1}g_{n-1}(\mathcal{E})+\mathcal{B}_ng_{n+1}(\mathcal{E})\right], \tag{21}$$

where $\mathcal{A}_n=\frac{a_n}{1+a_n}$ and $\mathcal{B}_n=1\Big/\sqrt{\left(1+a_n\right)\left(1+a_{n+1}\right)}$. This equivalent recursion relation is numerically preferred over (20) since in the limit as $n\to\infty$ the recursion coefficients $\mathcal{A}_n\to1$ and $\mathcal{B}_n\to0$, whereas in (20) the diagonal recursion coefficients diverge as $n^2$. In matrix notation, the recursion relation (21) is equivalent to the matrix eigenvalue equation $T\left|g\right\rangle=\frac{\mathcal{E}}{2\lambda}\left|g\right\rangle$, where *T* is the tridiagonal symmetric matrix whose elements are

$$T_{n,m}=\left(1-\frac{m}{\lambda}\right)\mathcal{A}_n\delta_{n,m}+\frac{V_0}{2\lambda}\left(\mathcal{B}_{n-1}\delta_{n,m+1}+\mathcal{B}_n\delta_{n,m-1}\right). \tag{22}$$

Taking $\lambda > m$ and for a given value of the potential strength $V_0$, Table 2 demonstrates the rapid convergence of the energy eigenvalues, $\mathcal{E}/2\lambda$, with the size of the matrix *T*. Figure 2 gives the relativistic energy spectrum as a function of the potential strength $V_0$. Note that the energy spectrum is independent of the sign of $V_0$. Normalizing the recursion relation (21) by taking $g_0(\mathcal{E}) = 1$,[‡] or equivalently factoring out $g_0(\mathcal{E})$ from the expansion of the wavefunction, it can be solved exactly for all coefficients $\{g_n(\mathcal{E})\}_{n=0}^{\infty}$ for a given choice of potential parameters $\{V_0, \lambda\}$ and energy eigenvalue $\mathcal{E}_k$, where $k = 0,1,2,...$ Figure 3 shows the spinor wavefunction components $\psi^{\pm}(\varepsilon_k, x)$ for the lowest few states with energy eigenvalues $\varepsilon_k = m + \mathcal{E}_k$ and for a given choice of potential parameters. These were calculated as follows

$$\psi^{\pm}(\varepsilon_k, x) = \sum_n f_n(\varepsilon_k)\phi_n^{\pm}(x) = \sum_{n=0}^{N} \sqrt{\frac{1+a_0}{1+a_n}}\, g_n(\mathcal{E}_k)\phi_n^{\pm}(x), \tag{23}$$

where *N* is the largest integer such that the sum (23) is stable at the desired accuracy. This value of *N* increases with the excitation number *k*. In Fig. 3 and with our choice of numerical routine, *N* is within the range 16 to 24. The number of nodes of $\psi^{+}(\varepsilon_k, x)$ and $\psi^{-}(\varepsilon_k, x)$ depends on the signs of $\mu$ and $\nu$. For $\mu = -\nu = \pm\frac{1}{2}$ they are the same and both equal to $k+1$. For $\mu = \nu = -\frac{1}{2}$, they are equal to $k+1$ and $k$ for $\psi^{+}(\varepsilon_k, x)$ and $\psi^{-}(\varepsilon_k, x)$, respectively. On the other hand, for $\mu = \nu = +\frac{1}{2}$ these are $k+1$ and $k+2$. However, the ground state ($k = 0$) may not obey this rule. Moreover, if we try to evaluate the wavefunction at an energy not equal to any of the eigenvalues $\{\varepsilon_k\}$ then the terms in the sum (23) will only produce an increasing number of oscillations with large amplitudes and will never achieve stable results. In fact, the sum of these oscillations for large *N* leads to destructive interference that should result in zero net value for the wave function.

## 4. Conclusion:

We succeeded in extending our nonrelativistic tridiagonal physics program to generate solvable potential problems to the relativistic 1D Dirac equation. The objective is to find exact solutions of the Dirac equation that do not belong to the already well-known class. The difficulty originating from the extra degree of freedom brought about by the two-component spinor wavefunction was alleviated by using the kinetic balance constraint that bounds the lower spinor component to the upper one. We started by writing the spinor wavefunction as a bounded infinite sum in a suitably chosen complete basis set whose parameters are chosen such that the matrix representation of the Dirac wave operator becomes tridiagonal and symmetric. This requirement transforms the wave equation into a symmetric three-term recursion relation for the expansion coefficients of the wavefunction. We solved the associated recursion relation and obtained the relativistic energy spectrum and corresponding state function. As an illustration, we considered the square well with half-sine potential bottom in the vector and scalar coupling mode under spin symmetry. We obtained the relativistic energy spectrum and showed that the nonrelativistic limit agrees with results found elsewhere.

---

‡ One can show that normalization of the spinor wavefunction gives $g_0(\mathcal{E}) \sim \sqrt{\rho(\mathcal{E})}$, where $\rho(\mathcal{E})$ is the weight function associated with the orthogonal polynomials solution of the three-term recursion relation (21). That is, $\int \rho(\mathcal{E}) P_n(\mathcal{E}) P_m(\mathcal{E}) d\mu(\mathcal{E}) = \delta_{nm}$, where $P_n(\mathcal{E}) = g_n(\mathcal{E})/g_0(\mathcal{E})$ and $d\mu(\mathcal{E})$ is some appropriate integration measure.

**Acknowledgements**:

The Authors would like to acknowledge the support provided by the Saudi Center for Theoretical Physics (SCTP) and King Fahd University of Petroleum and Minerals (KFUPM) during the progress of this work.

## Appendix A: The Jacobi basis

Using $y' = \lambda(1-y)^a(1+y)^b$, which gives $y'' = y'^2\left(\frac{b}{1+y} - \frac{a}{1-y}\right)$, and $\frac{d}{dy}\left|\phi_n^+\right\rangle = A_n(1-y)^\alpha$ $(1+y)^\beta\left(\frac{d}{dy} - \frac{\alpha}{1-y} + \frac{\beta}{1+y}\right)\left|P_n^{(\mu,\nu)}\right\rangle$, we can rewrite Eq. (7) as

$$\begin{aligned} J_{n,m} = &-\frac{A_n A_m}{\lambda}\left\langle P_n^{(\mu,\nu)}\right|(1-y)^{2\alpha}(1+y)^{2\beta}\left(1+\frac{Q_-}{2\lambda}\right)y'^2\left[\frac{d^2}{dy^2} + \left(\frac{2\beta+b}{1+y} - \frac{2\alpha+a}{1-y}\right)\frac{d}{dy}\right]\left|P_m^{(\mu,\nu)}\right\rangle \\ &-\frac{A_n A_m}{\lambda}\left\langle P_n^{(\mu,\nu)}\right|(1-y)^{2\alpha}(1+y)^{2\beta}y'\left(R - W + y'\frac{\dot{Q}_-}{2\lambda}\right)\frac{d}{dy}\left|P_m^{(\mu,\nu)}\right\rangle \\ &+\frac{A_n A_m}{\lambda}\left\langle P_n^{(\mu,\nu)}\right|(1-y)^{2\alpha}(1+y)^{2\beta}F(y)\left|P_m^{(\mu,\nu)}\right\rangle + n \leftrightarrow m \end{aligned} \quad \text{(A1)}$$

where,

$$\begin{aligned} F(y) = &\frac{\lambda Q_+}{2} - \left(R - W + y'\frac{\dot{Q}_-}{2\lambda}\right)\left[R + y'\left(\frac{\beta}{1+y} - \frac{\alpha}{1-y}\right)\right] \\ &+\left(1+\frac{Q_-}{2\lambda}\right)\left\{R^2 - y'\dot{R} - y'^2\left[\frac{\alpha(\alpha+a-1)}{(1-y)^2} + \frac{\beta(\beta+b-1)}{(1+y)^2} - \frac{\alpha b + \beta a + 2\alpha\beta}{1-y^2}\right]\right\} \end{aligned} \quad \text{(A2)}$$

Now, under coordinate transformation, the integral $\langle f|(...)|g\rangle = \lambda\int_{x_-}^{x_+} f(...)g\,dx$ gets mapped into $\lambda\int_{-1}^{+1} f(...)g\frac{dy}{y'}$. Moreover, the following four properties of the Jacobi polynomials will prove very useful

$$\left\{(1-y^2)\frac{d^2}{dy^2} - \left[(\mu+\nu+2)y + \mu - \nu\right]\frac{d}{dy} + n(n+\mu+\nu+1)\right\}P_n^{(\mu,\nu)}(y) = 0. \quad \text{(A3)}$$

$$(1-y^2)\frac{d}{dx}P_n^{(\mu,\nu)}(y) = -n\left(y + \frac{\nu-\mu}{2n+\mu+\nu}\right)P_n^{(\mu,\nu)}(y) + 2\frac{(n+\mu)(n+\nu)}{2n+\mu+\nu}P_{n-1}^{(\mu,\nu)}(y). \quad \text{(A4)}$$

$$A_n^2\int_{-1}^{+1}(1-y)^\mu(1+y)^\nu P_n^{(\mu,\nu)}(y)P_m^{(\mu,\nu)}(y)dx = \delta_{nm}. \quad \text{(A5)}$$

$$\begin{aligned} y\,P_n^{(\mu,\nu)}(y) = &\frac{\nu^2-\mu^2}{(2n+\mu+\nu)(2n+\mu+\nu+2)}P_n^{(\mu,\nu)}(y) \\ &+\frac{2(n+\mu)(n+\nu)}{(2n+\mu+\nu)(2n+\mu+\nu+1)}P_{n-1}^{(\mu,\nu)}(y) + \frac{2(n+1)(n+\mu+\nu+1)}{(2n+\mu+\nu+1)(2n+\mu+\nu+2)}P_{n+1}^{(\mu,\nu)}(y) \end{aligned} \quad \text{(A6)}$$

Imposing the differential equation (A3) in the first line of (A1) results in the following relations among the parameters

$$2\alpha + a = \mu + 1, \quad 2\beta + b = \nu + 1. \tag{A7}$$

On the other hand, using the differential property (A4) and orthogonality relation (A5) in the second line of Eq. (A1) dictate that

$$R - W + \tfrac{1}{2\lambda} V_-' = cy', \tag{A8}$$

where $c$ is a dimensionless real parameter. With these results, we can write (A1) as

$$J_{n,m} = 2c\lambda\,\Omega_{n,m} + 2\lambda A_n A_m \int_{-1}^{+1} (1-y)^\mu (1+y)^\nu G(y) P_n^{(\mu,\nu)}(y) P_m^{(\mu,\nu)}(y)\,dy, \tag{A9}$$

where

$$\begin{aligned} G(y) = {} & \frac{\lambda Q_+}{2}\frac{1-y^2}{y'^2} - c\left[\beta - \alpha - (\alpha+\beta)y + \frac{1-y^2}{y'}R\right] + \left(1 + \frac{Q_-}{2\lambda}\right)\Bigg\{\left(n + \tfrac{\mu+\nu+1}{2}\right)^2 \\ & - \frac{1}{4}\left(\mu^2 + \nu^2 + 2ab - 1\right) - \frac{1}{4}\left[\mu^2 - (a-1)^2\right]\frac{1+y}{1-y} - \frac{1}{4}\left[\nu^2 - (b-1)^2\right]\frac{1-y}{1+y} + \frac{1-y^2}{y'^2}\left(R^2 - y'\dot{R}\right)\Bigg\} \end{aligned} \tag{A10}$$

and $\Omega$ is the following tridiagonal symmetric matrix

$$\begin{aligned} \Omega_{n,m} = {} & (\nu - \mu)\frac{n(n+\mu+\nu+1)}{(2n+\mu+\nu)(2n+\mu+\nu+2)}\delta_{n,m} \\ & -(\mu+\nu+2)\left[\frac{1}{2n+\mu+\nu}\sqrt{\frac{n(n+\mu)(n+\nu)(n+\mu+\nu)}{(2n+\mu+\nu-1)(2n+\mu+\nu+1)}}\,\delta_{n,m+1}\right. \\ & \left. + \frac{1}{2n+\mu+\nu+2}\sqrt{\frac{(n+1)(n+\mu+1)(n+\nu+1)(n+\mu+\nu+1)}{(2n+\mu+\nu+1)(2n+\mu+\nu+3)}}\,\delta_{n,m-1}\right] \end{aligned} \tag{A11}$$

## Appendix B: The Laguerre Basis

For this problem, we choose $y \ge 0$, $y' = \lambda y^a e^{-by}$, where $a$ and $b$ are real dimensionless parameters, and take the following upper component of the discrete spinor basis

$$\phi_n^+(y) = A_n y^\alpha e^{-\beta y} L_n^\nu(y), \tag{B1}$$

where $L_n^\nu(y)$ is the Laguerre polynomial of order $n$ in $y$, $A_n = \sqrt{\frac{\Gamma(n+1)}{\Gamma(n+\nu+1)}}$, and the real parameters $\{\alpha,\beta,\nu\}$ will be determined later. The lower component $\phi_n^-(y)$ is obtained from the upper $\phi_n^+(y)$ by using the kinetic balance relation (4). Using $\frac{d}{dy}\left|\phi_n^+\right\rangle = A_n y^\alpha e^{-\beta y}\left(\frac{d}{dy} + \frac{\alpha}{y} - \beta\right)\left|L_n^\nu\right\rangle$ and $y'' = y'^2\left(\frac{a}{y} - b\right)$, the matrix representation of the wave operator (7) in the basis (B1) becomes

$$J_{n,m}=-\frac{A_nA_m}{\lambda}\left\langle L_n^\nu\left|y^{2\alpha}e^{-2\beta y}\left(1+\tfrac{Q_-}{2\lambda}\right)y'^2\left[\frac{d^2}{dy^2}+\left(\tfrac{2\alpha+a}{y}-2\beta-b\right)\frac{d}{dy}\right]\right|L_m^\nu\right\rangle$$
$$-\frac{A_nA_m}{\lambda}\left\langle L_n^\nu\left|y^{2\alpha}e^{-2\beta y}y'\left(R-W+y'\tfrac{\dot{Q}_-}{2\lambda}\right)\frac{d}{dy}\right|L_m^\nu\right\rangle \tag{B2}$$
$$+\frac{A_nA_m}{\lambda}\left\langle L_n^\nu\left|y^{2\alpha}e^{-2\beta y}F(y)\right|L_m^\nu\right\rangle+n\leftrightarrow m$$

where,

$$F(y)=\frac{\lambda Q_+}{2}-\left(R-W+y'\tfrac{\dot{Q}_-}{2\lambda}\right)\left[R+y'\left(\tfrac{\alpha}{y}-\beta\right)\right]$$
$$+\left(1+\tfrac{Q_-}{2\lambda}\right)\left\{R^2-y'\dot{R}-y'^2\left[\frac{\alpha(\alpha+a-1)}{y^2}-\frac{2\alpha\beta+\beta a+\alpha b}{y}+\beta(\beta+b)\right]\right\} \tag{B3}$$

In the first line of Eq. (B2), we impose the differential equation of the Laguerre polynomials,

$$\left[y\frac{d^2}{dy^2}+\left(\nu+1-y\right)\frac{d}{dy}+n\right]L_n^\nu(y)=0\,. \tag{B4}$$

This requires that $2\alpha+a=\nu+1$ and $2\beta+b=1$. On the other hand, we use the following differential property and orthogonality relation of the Laguerre polynomials in the second line of Eq. (B2)

$$y\frac{d}{dy}L_n^\nu(y)=nL_n^\nu(y)-(n+\nu)L_{n-1}^\nu(y)\,, \tag{B5}$$

$$A_n^2\int_0^\infty y^\nu e^{-y}L_n^\nu(y)L_m^\nu(y)dy=\delta_{nm}\,. \tag{B6}$$

These together with $y'=\lambda y^a e^{-by}$ dictate that $R-W+\frac{1}{2\lambda}V'_-=cy'$ so that a tridiagonal matrix representation is obtained, where $c$ is a dimensionless real parameter. Consequently, we can write (B2) as

$$J_{n,m}=c\lambda\left[-2n\,\delta_{n,m}+\sqrt{n\left(n+\nu\right)}\delta_{n,m+1}+\sqrt{\left(n+1\right)\left(n+\nu+1\right)}\delta_{n,m-1}\right]$$
$$+2\lambda A_nA_m\int_0^\infty y^\nu e^{-y}G(y)L_n^\nu(y)L_m^\nu(y)dy \tag{B7}$$

where

$$G(y)=\frac{\lambda Q_+}{2}\frac{y}{y'^2}-c\left(\alpha-\beta y+\tfrac{y}{y'}R\right)+\left(1+\tfrac{Q_-}{2\lambda}\right)\times$$
$$\left[\frac{y}{y'^2}\left(R^2-y'\dot{R}\right)+\left(n+\tfrac{\nu-ab+1}{2}\right)-\frac{1}{4}\frac{\nu^2-(a-1)^2}{y}+\frac{b^2-1}{4}y\right] \tag{B8}$$

Now, the orthogonality relation of the Laguerre polynomials (B6) and their recursion property,

$$y\,L_n^\nu=(2n+\nu+1)L_n^\nu-(n+\nu)L_{n-1}^\nu-(n+1)L_{n+1}^\nu\,, \tag{B9}$$

show that the matrix representation for the Dirac wave operator (B7) is tridiagonal if and only if $G(y)$ is a linear function of $y$.

As an example, we consider the spin symmetric coupling $S = V$. Therefore, the constraint below Eq. (B6) gives $R = W + cy'$. Now, if we also take $a = 1/2$ and $b = 0$ then we get $y(x) = \lambda^2 x^2/4$. Substituting all of that in (B8), we obtain

$$\begin{aligned} G(y) &= \frac{2V + m - \varepsilon}{2\lambda} - c\left(\alpha - \beta y + \tfrac{y}{y'} R\right) \\ &+ \left(1 - \tfrac{m+\varepsilon}{2\lambda}\right)\left[\frac{1}{\lambda^2}\left(R^2 - R'\right) + \left(n + \tfrac{\nu+1}{2}\right) - \frac{1}{4}\frac{\nu^2 - 1/4}{y} - \frac{1}{4} y\right] \end{aligned} \quad \text{(B10)}$$

One way to make this a linear function in $y$ is to take $\frac{R^2 - R'}{\lambda^2} = \frac{\nu^2 - 1/4}{4y} + \xi y + \eta$ subject to the constraint that also $\frac{y}{y'} R$ is a linear function in $y$, where $\xi$ and $\eta$ are dimensionless real constants. Additionally, we should also take $V(y) = V_0 y$, which is the harmonic oscillator potential. A particular solution of the nonlinear equation for $R$, which is suggested by the requirement that $\frac{y}{y'} R$ be a linear function in $y$, is $R(x) = \frac{\gamma}{x} + \frac{1}{2}\hat{c}\lambda^2 x$ giving $\nu^2 = \left(\gamma + \frac{1}{2}\right)^2$, $\xi = \hat{c}^2$ and $\eta = \hat{c}\left(\gamma - \frac{1}{2}\right)$. Moreover, the pseudo-scalar potential function becomes $W(x) = \frac{\gamma}{x} + \frac{1}{2}\tau\lambda^2 x$, where $\hat{c} - c = \tau$. Therefore, the set of physical parameters of the problem is $\{V_0, \gamma, \tau, \lambda\}$ and thus only the difference between the two parameters $c$ and $\hat{c}$ has physical significance. Consequently, we can set one of them to zero, and to simplify the computation we choose $c$ to vanish giving

$$G(y) = \frac{V_0}{\lambda} y + \frac{m - \varepsilon}{2\lambda} + \left(1 - \tfrac{m+\varepsilon}{2\lambda}\right)\left[\left(\tau^2 - \frac{1}{4}\right) y + \tau\left(\gamma - \frac{1}{2}\right) + \left(n + \tfrac{\nu+1}{2}\right)\right]. \quad \text{(B11)}$$

Finally, using the recursion relation and orthogonal property of the Laguerre polynomials, we obtain the following tridiagonal matrix representation of the Dirac wave operator (B7)

$$\begin{aligned} J_{n,m} &= \left\{m - \varepsilon + \lambda\tau(2\gamma - 1)\left(1 - \tfrac{m+\varepsilon}{2\lambda}\right) + \left(n + \tfrac{\nu+1}{2}\right)\left[4V_0 + \lambda\left(4\tau^2 + 1\right)\left(1 - \tfrac{m+\varepsilon}{2\lambda}\right)\right]\right\}\delta_{n,m} \\ &- \frac{1}{2}\left[4V_0 + \lambda\left(4\tau^2 - 1\right)\left(1 - \tfrac{m+\varepsilon}{2\lambda}\right)\right]\left[\sqrt{n(n+\nu)}\,\delta_{n,m+1} + \sqrt{(n+1)(n+\nu+1)}\,\delta_{n,m-1}\right] \end{aligned} \quad \text{(B12)}$$

Therefore, the diagonal representation is obtained if $V_0 = 0$ and $\tau = 1/2$. In that case, the energy spectrum formula reads as follows

$$\frac{\varepsilon - m}{2\lambda} \Big/ \left(1 - \tfrac{\varepsilon+m}{2\lambda}\right) = n + \frac{1}{2}\left(\nu + \gamma + \frac{1}{2}\right) = \begin{cases} n + \gamma + \frac{1}{2} & ,\nu = \gamma + \frac{1}{2} \\ n & ,\nu = -\gamma - \frac{1}{2} \end{cases}. \quad \text{(B13)}$$

Taking the nonrelativistic limit, $\varepsilon \approx m + E$ where $E << m$, gives

$$E_n = 2(\lambda - m)\left[n + \frac{1}{2}\left(\nu + \gamma + \frac{1}{2}\right)\right], \quad \text{(B14)}$$

which agrees with the nonrelativistic result under the parameter map $\lambda \to m + (\lambda/2)^2$ and with $\gamma$ being an effective angular momentum quantum number. Expanding the spinor wavefunction as $|\psi\rangle = \sum_n f_n |\phi_n\rangle$ and substituting in the wave equation $J|\psi\rangle = 0$, we obtain with the use of (B12) the following three-term recursion relation for the expansion coefficients $\{f_n\}$

$$\left[\frac{2\mathcal{E}}{\lambda}-2\tau(2\gamma-1)\left(1-\tfrac{\varepsilon+m}{2\lambda}\right)\right]f_n=\left[\frac{4V_0}{\lambda}+\left(4\tau^2+1\right)\left(1-\tfrac{\varepsilon+m}{2\lambda}\right)\right]a_nf_n \\ -\left[\frac{4V_0}{\lambda}+\left(4\tau^2-1\right)\left(1-\tfrac{\varepsilon+m}{2\lambda}\right)\right]\left(b_{n-1}f_{n-1}+b_nf_{n+1}\right), \quad \text{(B15)}$$

where $\mathcal{E}=\varepsilon-m$, $a_n=2n+\nu+1$, and $b_n=\sqrt{(n+1)(n+\nu+1)}$. We simplify the problem by choosing $\tau=0$ resulting in one-parameter potentials, $V(y)=V_0y$ and $W(x)=\gamma/x$. Dividing (B15) by $\frac{4V_0}{\lambda}-\left(1-\frac{\varepsilon+m}{2\lambda}\right)$, this could be written as

$$\left(z\sin\theta\right)f_n=\left[\left(2n+\nu+1\right)\cos\theta\right]f_n-\sqrt{n(n+\nu)}f_{n+1}-\sqrt{(n+1)(n+\nu+1)}f_{n+1}, \quad \text{(B16)}$$

where $\cos\theta=\frac{4V_0-\left(m-\lambda+\mathcal{E}/2\right)}{4V_0+\left(m-\lambda+\mathcal{E}/2\right)}$ and $z=\mathcal{E}/2\sqrt{V_0\left(m-\lambda+\mathcal{E}/2\right)}$ with $\mathcal{E}\ge 2(\lambda-m)$ for $V_0>0$ so that $0\le\theta\le\pi$. Comparing this with the recursion relation of the normalized version of the Meixner-Pollaczek polynomials that reads [14]

$$\left(x\sin\theta\right)P_n^{\mu}(x;\theta)=-\left[\left(n+\mu\right)\cos\theta\right]P_n^{\mu}(x;\theta) \\ +\tfrac{1}{2}\sqrt{n(n+2\mu-1)}P_{n-1}^{\mu}(x;\theta)+\tfrac{1}{2}\sqrt{(n+1)(n+2\mu)}P_{n+1}^{\mu}(x;\theta) \quad \text{(B17)}$$

we conclude that $f_n(\varepsilon)=P_n^{\frac{\nu+1}{2}}(-z/2;\theta)$.

Mathematics and Informatics, Number 98-17 (Delft University of Technology, Delft, 1998) page 37

**Table Captions:**

**Table 1**: The upper and lower components of the Jacobi spinor basis for $a = b = 1/2$ and associated with the four cases $\mu = \nu = \pm\frac{1}{2}$ and $\mu = -\nu = \pm\frac{1}{2}$. The value of the basis at the two boundaries of configuration space $(-L/2, +L/2)$ is given where $\times$ indicates non-zero value.

**Table 2**: The energy eigenvalues, $\mathcal{E}/2\lambda$, associated with the lowest energy states. The table demonstrates the rapid convergence with the size of the basis. We took $m = 1$, $\lambda = 1.2$, $V_0 = 0.5$, and $\mu = \nu = -\frac{1}{2}$.

**Figure Captions:**

**Fig. 1**: Plot of the spinor basis given in Table 1 for $n$ = 6 showing the oscillatory behavior and the typical (almost $\pi/2$) phase shift between the upper (solid curve) and lower (dashed curve) components. The horizontal axis is the configuration space coordinate $x \in [-\frac{\pi}{\lambda}, +\frac{\pi}{\lambda}] = [-\frac{1}{2}L, +\frac{1}{2}L]$. We took (a) $\mu = \nu = -\frac{1}{2}$, (b) $\mu = \nu = +\frac{1}{2}$, (c) $\mu = -\nu = -\frac{1}{2}$, (d) $\mu = -\nu = +\frac{1}{2}$.

**Fig. 2**: The relativistic energy spectrum as a function of the potential strength $V_0$, which is independent of its sign. The traces from down up correspond to the ground state, first, second, and third excited state, respectively. We took $m = 1$ and $\lambda = 1.5$ for (a) $\mu = \nu = -\frac{1}{2}$, (b) $\mu = -\nu = \pm\frac{1}{2}$, (c) $\mu = \nu = +\frac{1}{2}$. Note the degeneracy in case (b).

**Fig. 3**: The un-normalized components of the eigen-energy spinor as given by Eq. (23) for few of the lowest excited states of Table 2. The solid (dashed) curve corresponds to the upper (lower) component. The horizontal axis is the configuration space coordinate $x \in [-\frac{1}{2}L, +\frac{1}{2}L]$.

**Table 1**

| $(\mu,\nu)$ | $\phi_n^+(x)$ | $\phi_n^-(x)=\sqrt{1-y^2}\,\dfrac{d\phi_n^+}{dy}$ | $\phi_n^+(\pm\frac{1}{2}L)$ | $\phi_n^-(\pm\frac{1}{2}L)$ |
|---|---|---|---|---|
| $(-1/2,-1/2)$ | $\sqrt{\frac{2}{\pi}}T_n(\sin\lambda x)$ | $n\sqrt{\frac{2}{\pi}}(\cos\lambda x)U_{n-1}(\sin\lambda x)$ | (×,×) | (0,0) |
| $(+1/2,+1/2)$ | $\sqrt{\frac{2}{\pi}}(\cos\lambda x)U_n(\sin\lambda x)$ | $-(n+1)\sqrt{\frac{2}{\pi}}T_{n+1}(\sin\lambda x)$ | (0,0) | (×,×) |
| $(-1/2,+1/2)$ | $\frac{1}{\sqrt{\pi}}\sqrt{1+\sin(\lambda x)}\,V_n(\sin\lambda x)$ | $\frac{n+1/2}{\sqrt{\pi}}\sqrt{1-\sin(\lambda x)}\,W_n(\sin\lambda x)$ | (0,×) | (×,0) |
| $(+1/2,-1/2)$ | $\frac{1}{\sqrt{\pi}}\sqrt{1-\sin(\lambda x)}\,W_n(\sin\lambda x)$ | $-\frac{n+1/2}{\sqrt{\pi}}\sqrt{1+\sin(\lambda x)}\,V_n(\sin\lambda x)$ | (×,0) | (0,×) |

**Table 2**

| $k$ | 10×10 | 15×15 | 20×20 | 30×30 |
|---|---|---|---|---|
| 0 | –0.1180273824 | –0.1180273824 | –0.1180273824 | –0.1180273824 |
| 1 | 0.0957203979 | 0.0957203979 | 0.0957203979 | 0.0957203979 |
| 2 | 0.1410291052 | 0.1410291051 | 0.1410291051 | 0.1410291051 |
| 3 | 0.1540674139 | 0.1540664762 | 0.1540664762 | 0.1540664762 |
| 4 | 0.1594092708 | 0.1592706381 | 0.1592706381 | 0.1592706381 |
| 5 | 0.1631558065 | 0.1618247923 | 0.1618246774 | 0.1618246774 |
| 6 | 0.1668739393 | 0.1632721765 | 0.1632573337 | 0.1632573337 |
| 7 | 0.1700465453 | 0.1643388284 | 0.1641386243 | 0.1641386049 |
| 8 | 0.1764243575 | 0.1654457643 | 0.1647206731 | 0.1647183588 |
| 9 | 0.2293185740 | 0.1665645929 | 0.1651610351 | 0.1651197049 |

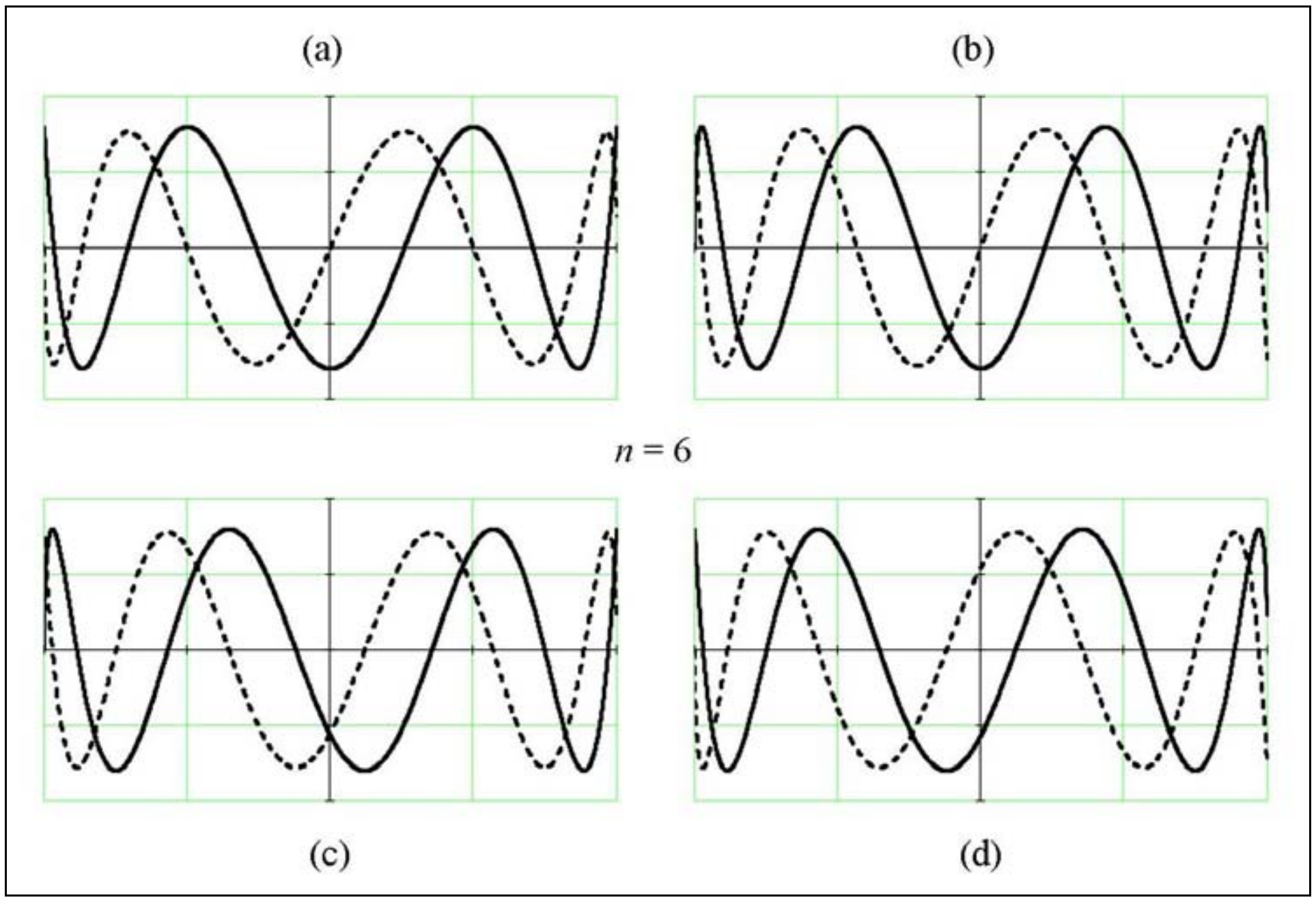

**Fig. 1**

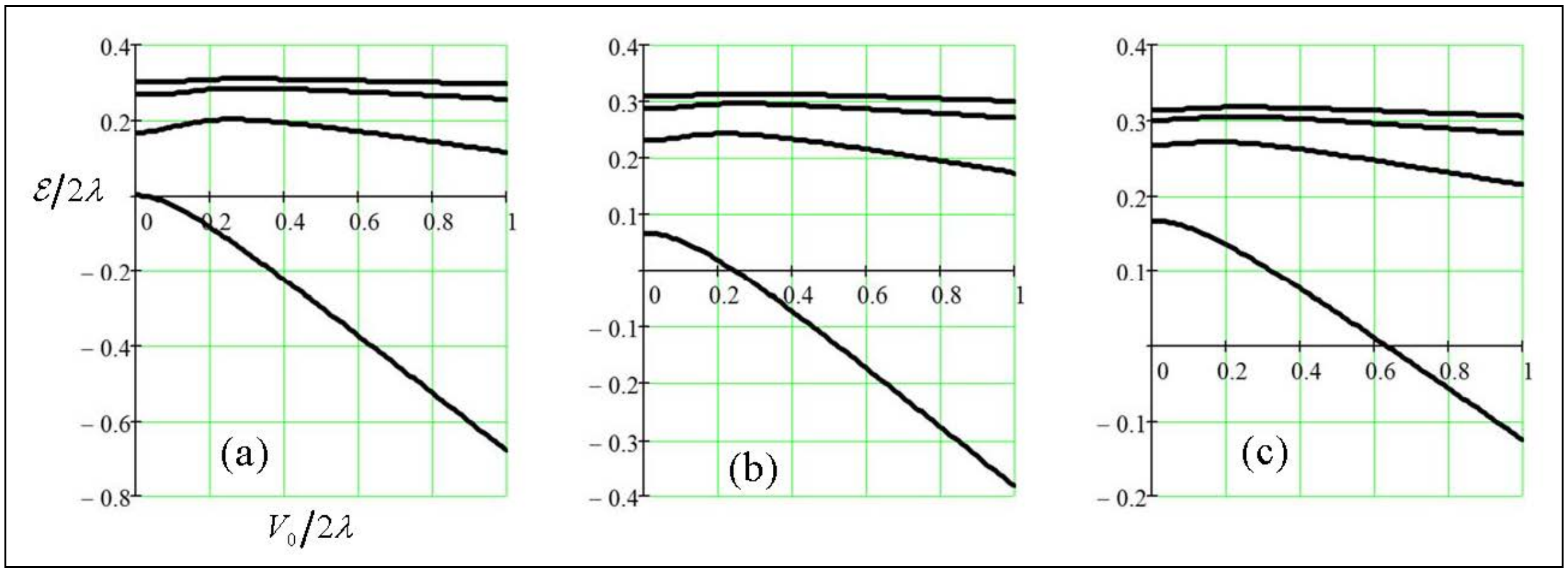

**Fig. 2**

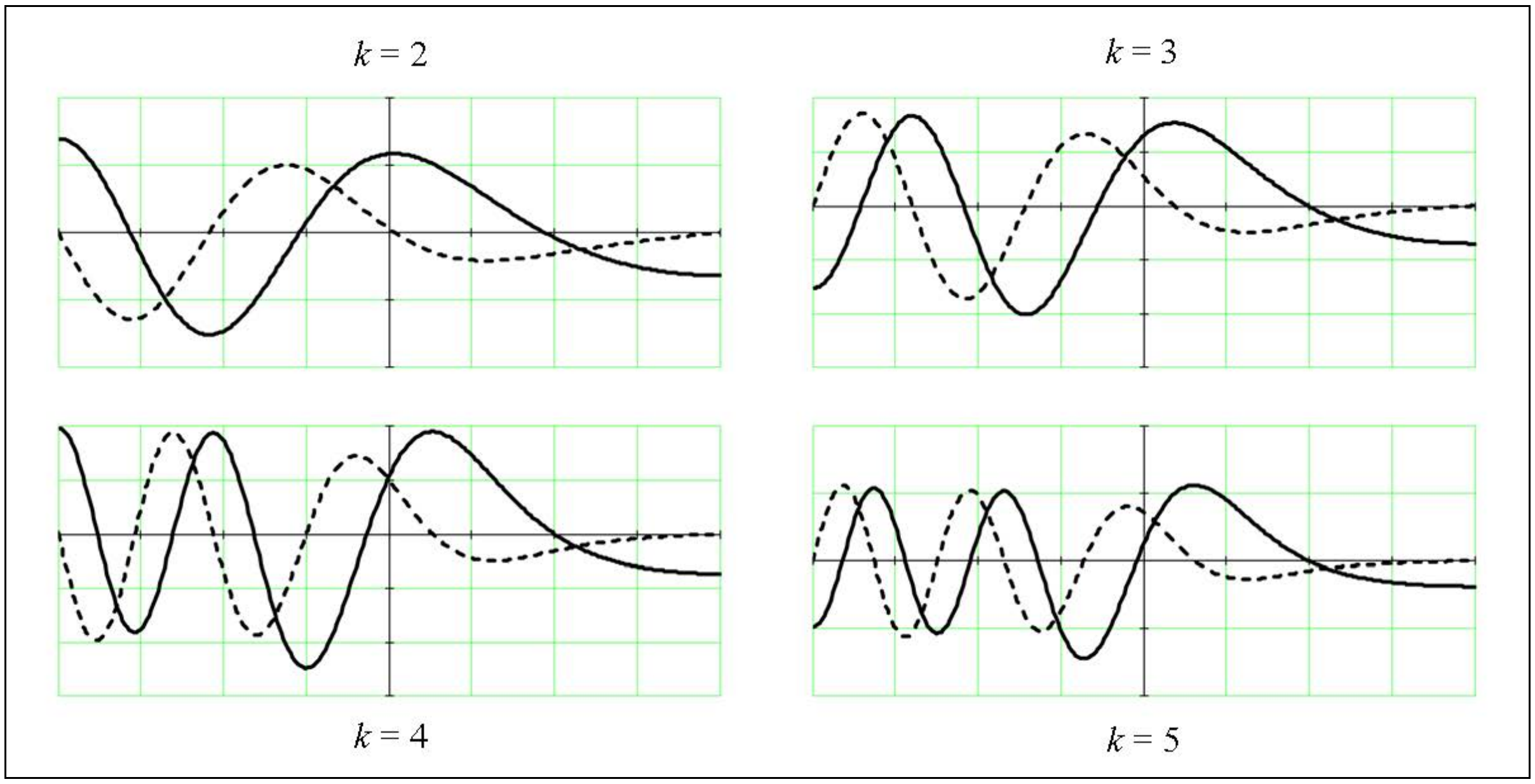

**Fig. 3**